\documentclass[amsmath,amssymb,prl,twocolumn,superscriptaddress]{revtex4-1}

\usepackage[utf8]{inputenc}
\usepackage{graphicx}
\usepackage{dcolumn}
\usepackage{bm}
\usepackage{siunitx}
\usepackage[version=4]{mhchem}
\usepackage{natbib}
\usepackage{float}

\usepackage[shortlabels]{enumitem}
\usepackage{lipsum}
\usepackage{xcolor}
\usepackage{mathtools}
\usepackage{braket}
\usepackage{amsmath}%
\usepackage{MnSymbol}%
\usepackage{wasysym}%
\usepackage{cancel}
\usepackage{physics}
\usepackage{caption}
\usepackage{subcaption}
\usepackage[english]{babel}
\usepackage[T1]{fontenc}
\usepackage{placeins}
\usepackage{comment}

\usepackage[colorlinks,citecolor=blue,urlcolor=red]{hyperref}
\usepackage{cleveref} 
\usepackage{chemformula}

\makeatletter
\def\maketitle{
\@author@finish
\title@column\titleblock@produce
\suppressfloats[t]}
\makeatother

\begin{document}

\title{Coherence of a non-equilibrium polariton condensate across the interaction-mediated phase transition}

\author{P. Comaron}
\thanks{These two authors contributed equally to this work}
\email{p.comaron@ucl.ac.uk}
\address{Department of Physics and Astronomy, University College London,
Gower Street, London, WC1E 6BT, United Kingdom} 

\author{E. Estrecho}
\thanks{These two authors contributed equally to this work}
\email{eliezer.estrecho@anu.edu.au}
\affiliation{ARC Centre of Excellence in Future Low-Energy Electronics Technologies and Department of Quantum Science and Technology, Research School of Physics, The Australian National University, Canberra, ACT 2601, Australia}

\author{M. Wurdack}
\thanks{current address: Hansen Experimental Physics Laboratory and Department of Neurosurgery, Stanford University, Stanford, CA 94305}
\affiliation{ARC Centre of Excellence in Future Low-Energy Electronics Technologies and Department of Quantum Science and Technology, Research School of Physics, The Australian National University, Canberra, ACT 2601, Australia}

\author{M. Pieczarka}
\affiliation{ARC Centre of Excellence in Future Low-Energy Electronics Technologies and Department of Quantum Science and Technology, Research School of Physics, The Australian National University, Canberra, ACT 2601, Australia}
\affiliation{Department of Experimental Physics, Faculty of Fundamental Problems of Technology, Wroclaw University of Science and Technology, Wyb. Wyspia\'{n}skiego 27, 50-370 Wroc\l aw Poland}

\author{M.~Steger}
\thanks{current address: National Renewable Energy Laboratory, Golden, CO 80401, USA}
\affiliation{Department of Physics and Astronomy, University of Pittsburgh, Pittsburgh, PA 15260, USA}%

\author{D.~W.~Snoke}
\affiliation{Department of Physics and Astronomy, University of Pittsburgh, Pittsburgh, PA 15260, USA}%

\author{K.~West}
\affiliation{Department of Electrical Engineering, Princeton University, Princeton, NJ 08544, USA}%

\author{L.~N.~Pfeiffer}
\affiliation{Department of Electrical Engineering, Princeton University, Princeton, NJ 08544, USA}

\author{A. G. Truscott}
\affiliation{Department of Quantum Science and Technology, Research School of Physics, The Australian National University, Canberra, ACT 2601, Australia}

\author{M. Matuszewski}
\affiliation{Institute of Physics, Polish Academy of Sciences, Al. Lotnikow 32/46, 02-668 Warsaw, Poland}
\affiliation{Center for Theoretical Physics, Polish Academy of Sciences, Al. Lotnikow 32/46, 02-668 Warsaw, Poland}

\author{M. Szymanska}
\email{m.szymanska@ucl.ac.uk}
\address{Department of Physics and Astronomy, University College London,
Gower Street, London, WC1E 6BT, United Kingdom} 

\author{E. A. Ostrovskaya}
\email{elena.ostrovskaya@anu.edu.au}
\affiliation{ARC Centre of Excellence in Future Low-Energy Electronics Technologies and Department of Quantum Science and Technology, Research School of Physics, The Australian National University, Canberra, ACT 2601, Australia}


\begin{abstract}
The emergence of spatial coherence in a confined two-dimensional Bose gas of exciton polaritons with tuneable interactions offers a unique opportunity to explore the role of interactions in a phase transition in a driven-dissipative quantum system, where both the phase transition and thermalisation are mediated by interactions. 
We investigate, experimentally and numerically, the  phase correlations and steady-state properties of the gas over a wide range of interaction strengths by varying the photonic/excitonic fraction of the polaritons and their density. We find that the first order spatial coherence function exhibits algebraic decay consistent with the Berezinskii–Kosterlitz–Thouless (BKT) phase transition. 
Surprisingly, the exponent of the algebraic decay is inversely proportional to the coherent density of polaritons, in analogy to equilibrium superfluids above the BKT transition, but with a different proportionality constant.
Our work paves the way for future investigations of the phenomenon of phase transitions and superfluidity in a driven-dissipative setting.
\end{abstract}

\maketitle


\noindent\textbf{\large {Introduction} \normalsize}

The physics of collective phenomena in quantum many-body systems is strongly connected to dimensionality, interactions and to the intrinsic equilibrium/nonequilibrium nature of the system. 
A gas of identical bosons is a perfect example of the crucial role that these factors play in the onset of order~\cite{Hohenberg1967}. The three-dimensional (3D) gas of bosons can undergo a Bose–Einstein condensation (BEC), which results in a state of matter exhibiting long-range order. While thermal fluctuation preclude the formation of the long-range order in a two-dimensional (2D) gas, it can instead undergo the Berezinskii–Kosterlitz–Thouless (BKT) transition and exhibit superfluidity below a finite critical temperature. {Within the superfluid phase, the long-range decay exponent of the spatial correlations  of the system is predicted to be a function of temperature and the superfluid density \cite{Berezinskii1971,Berezinskii1973}}. 
In order to achieve a BEC phase transition in 2D, the bosons have to be non-interacting and trapped \cite{Dalfovo1999}. Both BEC and BKT transitions have been explored in the system of cold atomic gases by tuning the system's dimensionality or the interaction strength using Feshbach resonances \cite{Hadzibabic_2008,Fletcher2015,chomaz2015,Simula2005transition,bisset2009,HadzibabicDalibard2011,Nazarenko2014,Comaron2019}. 

2D bosonic systems, such as cavity photons \cite{klaers2010bose} or exciton-polaritons ~\cite{byrnes2014exciton, carusotto2013quantum}, represent a class of systems with driving and dissipation, where the full thermalisation (energy relaxation) may not be achieved ~\cite{deng2010exciton} and the physics is inherently nonequilibrium. While cavity photons closely resemble an “ideal” gas, exciton polaritons (polaritons herein) arise from the strong coupling of cavity photons and excitons in a semiconductor and interact strongly due to their excitonic component. Theoretical~\cite{comaron2020BKT,Mei2021,Gladlin2019BKT,Gladlin2023BKT}, and experimental~\cite{nitsche2014algebraic,caputo2018} investigations demonstrated the relevance of the interaction-dominated BKT physics to polariton systems under nonresonant pumping, when the coherence in the system develops spontaneously. 
{
However, so far it remains unclear how the coherence in the intrinsically driven-dissipative polariton system depends on the interactions, how it is affected by confinement, and whether there exists, in analogy with conservative systems, any relation between the decay of coherence and the superfluid fraction. These questions are also interesting because both the thermalisation (energy relaxation) and the phase transition in a polariton system are driven by interactions \cite{deng2010exciton}.
}

In this work, we investigate the phase coherence, i.e. the first-order spatial correlation function, of a confined polariton condensate in an optical microcavity. The confinement allows us to explore the coherence features at high densities, when the interaction energy dominates over the kinetic energy. This is in contrast to the previous work, where the signatures of the BKT transition were detected by studying the coherence of a low-density polariton flow ~\cite{caputo2018}. In particular, we investigate the coherence behaviour while varying the polariton interaction energy, which is proportional to the polariton density and the strength of interparticle interactions. The variation of the interaction strength is achieved by changing the energy detuning between the cavity photon and the exciton, which changes the excitonic fraction of the polariton. The larger excitonic fraction corresponds to stronger interparticle interaction. In turn, the condensate density is controlled by the power of the optical pump.

Our study reveals that the trapped, condensed polaritons exhibit different regimes of spatial coherence depending on the interaction strength. Full coherence across the spatial extent of the system is reached for stronger interacting (more excitonic) polaritons. The study of coherence dependence on the density for a fixed interaction strength (fixed excitonic fraction) shows similar behaviour, as the interaction energy grows with increasing density. However, at very high excitonic fractions and pump powers the coherence decreases due an increase in various decoherence mechanisms. Importantly, in the limit of weakly interacting (more photonic) polaritons, full thermalisation does not occur, and multi-mode condensation takes place instead of a ground state BEC predicted for a non-interacting trapped 2D bosonic system \cite{Dalfovo1999}. This confirms that the thermalisation in this system is driven by interactions and is inhibited in the photonic regime, making it difficult to reach the weakly or non-interacting BEC state. Since both condensation and thermalisation processes are strongly interaction-driven, observation in polariton condensates  of a clean crossover (pioneered in cold atoms using Feshbach resonances \cite{Chin2010} from the interacting BKT to the non-interacting 2D BEC regimes by tuning the interaction strength is challenging. 

{To further characterise the transition to the BKT phase in this driven-disspative system, we introduce the concept of “coherent fraction” of the polariton condensate, and study the relation to the power-law exponent, describing the decay of the spatial first-order correlation function. Our analysis demonstrates growth and decay of the coherent fraction with the increasing interaction energy, consistent with the non-monotonic behaviour of the first-order spatial correlation function.
Surprisingly, our findings also suggest a clear  relationship between the power-law exponent, the coherent fraction and the effective temperature of a polariton system, in analogy with the relationship between the power-law exponent and superfluid fraction for quantum gases in thermal equilibrium \cite{ProkofevSvistunov2002}.}

\begin{figure}
\includegraphics[width=\linewidth]{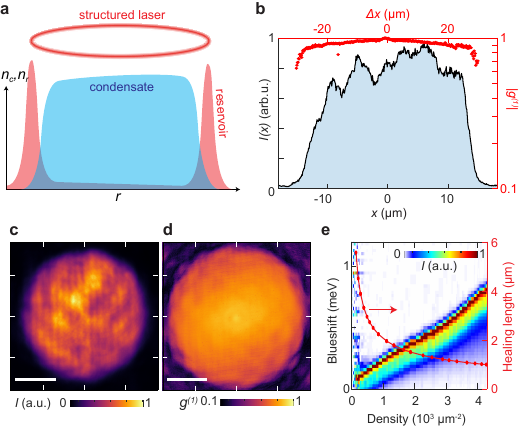}
\caption{\textbf{Experimental configuration}. {\bf a}, Sketch of the excitation scheme where a ring-shaped laser excites the sample and creates a ring-shaped reservoir of excitons with the density $n_r$ that traps the top-hat condensate with the density $n_c$. {{\bf b}, Comparison of the cross-sections of the measured intensity of the condensate PL $I$ (proportional to the polariton density) and the first-order correlation function $|g^{(1)}|$ showing that the spatial extent of $|g^{(1)}|$ is limited by the size of the condensate. Scale bar: 10~$\mu$m. {\bf c}, Example experimental image of the real-space distribution of the condensate. Scale bar: 20~$\mu$m. {\bf d}, Measured 2D $g^{(1)}(\mathbf{r})$ of the condensate in {\bf c}, acquired using a modified Michelson interferometer (see `Methods'). {\bf e}, Example density dependence of the blueshift and healing length. The color represents the normalized spectra for each density datapoint. The excitonic fraction here is $|X|^2 = 0.34$.}}
\label{fig:fig1}
\end{figure}

\begin{figure*}[htp]
\centering
\includegraphics[width=\textwidth]{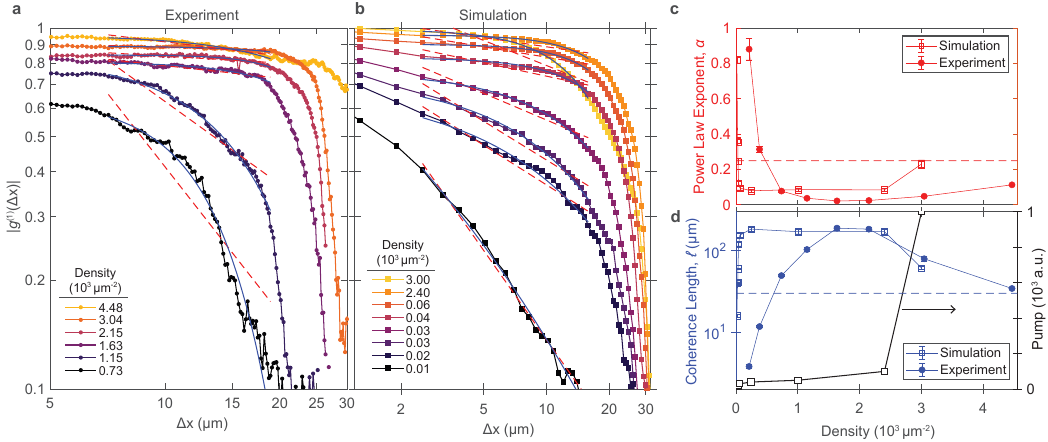}
\caption{\textbf{Density dependence of $g^{(1)}(\Delta x)$}. {\bf a}, Logarithmic plots of $g^{(1)}$ at {$|X|^2=0.37$} for different densities. Dots are experimental data and solid lines are the power-law fits (see text). {\bf b} $g^{(1)}(\Delta x)$ calculated using the stochastic model \eqref{eq:wigner} with the experimental parameters. Extracted {\bf c} power-law exponents $\alpha$ and {\bf d} coherence lengths $\ell$, respectively, as a function of density for different excitonic fractions $|X|^2$.
In \textbf{c} and \textbf{d} we report as a dashed line the equilibrium limit for the algebraic exponent $\alpha=0.25$~\cite{Berezinskii1973} and the box size, respectively.
}
\label{fig:fig2}
\end{figure*}

\

\noindent\textbf{\large {Results} \normalsize}

We study polariton condensates confined in an optically-induced circular trap~\cite{estrecho2019direct, askitopoulos2013polariton, pieczarka2020observation, sun2017, sun2017direct,Alnatah2024} in the steady-state regime. An off-resonant quasi-continuous wave (cw) excitation laser, shaped into a ring profile, pumps the microcavity sample with embedded GaAs quantum wells (see Methods), creating a similarly shaped excitonic reservoir that feeds and repels polaritons. As schematically shown in Fig.~\ref{fig:fig1}(a), above a critical pump power (reservoir density), this configuration produces a trapped condensate inside the ring. The constant replenishment of the reservoir by the excitation laser ensures that the condensate density is relatively constant despite the losses due to the short polariton lifetime ($\sim$100 ps). The polaritons decay by emitting photons, which we detect in the experiment, and the intensity of the photoluminescence (PL) is directly proportional to the polariton density [Fig.~\ref{fig:fig1}(b)]. The  real space image [Fig.~\ref{fig:fig1}(c)], momentum space distribution, spectra, and first-order correlation functions of the PL are used to infer the information about the polariton condensate.

At high densities of the condensate, when its interaction energy dominates over the kinetic energy, the density profile shows a relatively flat-top distribution [Fig.~\ref{fig:fig1}(b)]. By virtue of the Thomas-Fermi approximation, this reflects the flat, box-like profile of the underlying trapping potential~\cite{estrecho2019direct, pieczarka2020observation}. The uniform profile of both the condensate and the reservoir-induced potential ensures that the densities of the condensate and the reservoir are spatially homogeneous near the centre of the trap.
Note that this regime only works in our experiments for sufficiently large traps with a diameter around $30~\mu$m. Smaller traps lead to quasi-harmonic potentials while very large traps need high excitation powers, beyond our experimental capability, to reach the high-density regime. {In this work, we probe interaction energies up to 700~$\mu\mathrm{eV}$ (see SI)}.

A wedge in the microcavity sample, arising during the fabrication process, leads to a spatial variation of the exciton-photon detuning~\cite{sun2017direct, estrecho2019direct}, which translates to the variation of the excitonic fraction quantified by the Hopfield coefficient $0\leq |X|^2\leq 1$ (see 'Methods'). The range of the excitonic Hopfield coefficient accessible in our sample is $0.21 \leq |X|^2\leq 0.6$. The excitonic fraction in the polariton quasiparticle is varied within this range by moving the laser excitation spot across the sample.

We use a modified Michelson interferometer~\cite{caputo2018topological} to measure the two-dimensional first-order correlation function |$g^{(1)}(\Delta x, \Delta y)$| (see 'Methods'), where $(\Delta x,\Delta y)$ is the displacement from the autocorrelation point.
An example of $|g^{(1)}|$ is shown in Fig.~\ref{fig:fig1}(b) for an excitonic fraction of $|X|^2=0.37$ with the peak density of $n=4.13\times10^3\mu\mathrm{m}^{-2}$. It features a peak at the autocorrelation point $\Delta\mathbf{r}=0$, which characterises short-range correlations. The slowly decaying shoulder at larger distances characterises the long-range order in this system which is the main interest of this work. By comparing the $|g^{(1)}|$ to the condensate profile, one can clearly see in Figs.~\ref{fig:fig1}(c),(d) that the long-range order is limited only by the size of the condensate. Note that despite the small inhomogeneities of the measured condensate profile, which can arise from both imperfections of the imaging setup and sample disorder, the measured $|g^{(1)}|$ remains smooth, which strongly suggests that our measurement is not affected by the small perturbations of the steady-state density.

\

\begin{figure*}[htp]
\centering
\includegraphics[width=\textwidth]{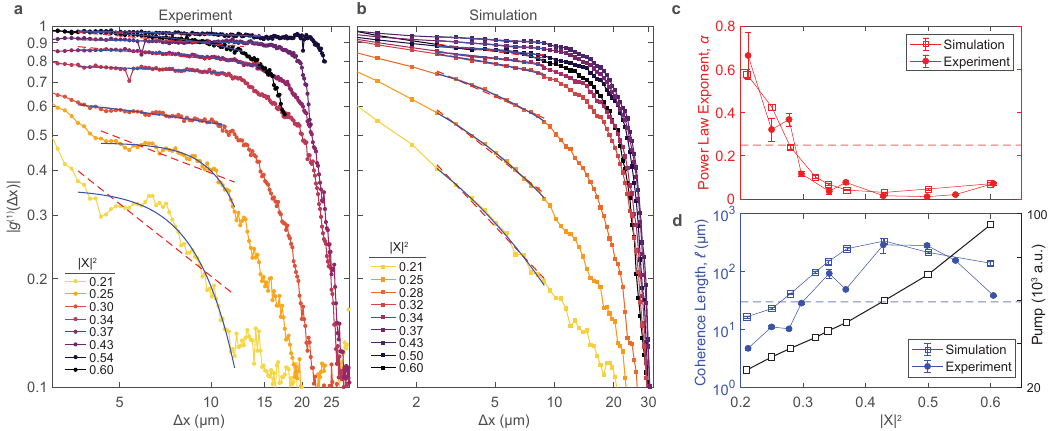}
\caption{\textbf{Excitonic fraction dependence of $g^{(1)}(\Delta x)$}. {\bf a}, Logarithmic plots of $g^{(1)}$ at constant density $n\approx1\times 10^3 \mu\mathrm{m}^{-2}$ for different excitonic fractions $|X|^2$. Dots are experimental data and solid lines are the power-law fits. Theoretical results are reported in panel {\bf b}.  
{Extracted {\bf c} power-law exponents $\alpha$ and {\bf d} coherence lengths $\ell$ as a function of the excitonic fraction. Experimental data (solid blue line and blue filled circles) compared to the simulation data (solid blue line and blue bare squares). The pump strength in simulations is shown by a black solid line and squares.}
}
\label{fig:fig3}
\end{figure*}

\noindent\textbf{{Density dependence of coherence at a fixed excitonic fraction.} --}
To understand how the condensate interaction energy affects the first-order correlation function, we performed a series of measurements at different excitation powers above the condensation threshold, resulting in different condensate densities $n$. Here, $n$ is the average density in a small area around the centre of the condensate.

It is important to highlight that the size of the condensate (diameter of $\approx30~\mu m$) is an order of magnitude larger than the healing length $\xi=\hbar/\sqrt{2gnm}$, where $g$ is the polariton-polariton interaction strength~\cite{estrecho2019direct}, $m$ is the polariton mass, and $\hbar$ is the Planck's constant. The healing length ranges from 6~$\mu$m down to 1~$\mu$m at the probed densities, while the measured condensate blueshift {can reach $800\mu\mathrm{eV}$ as shown in Fig.~\ref{fig:fig1}(e). The condensate interaction energy $gn$ is less than this due to nonvanishing reservoir density inside the trap~\cite{pieczarka2020observation, estrecho2021low, pieczarka2022bogoliubov} (see SI)}. The small healing lengths relative to the system size guarantee that we are indeed probing the effect of interactions on the long-range correlations of the condensate. If $\xi\sim D$, the long-range universal behaviour in $g^{(1)}$ is masked by other short-range effects.

Figure~\ref{fig:fig2}a presents the measured $|g^{(1)}|$ at a fixed value of the excitonic fraction {$|X|^2=0.37$}, which clearly shows that the long-range coherence increases with density. At low density, $|g^{(1)}(\Delta x)|$ quickly decays with distance $\Delta x$. This spatial decay decreases with increasing density until  $|g^{(1)}|$ becomes almost flat.

To quantify the effect of interactions on the coherence length, we fit the experimental data using the functions~\cite{Dalfovo1999}
\begin{equation}
g^{(1)}_\text{exp}(r) \propto e^{-r/\ell}  \quad \text{and} \quad g^{(1)}_\text{alg}(r) \propto r^{-\alpha} \, .
\label{eq:gonefitrel}
\end{equation}
According to the BKT theory of a spatially isotropic condensate, the fast, exponential (slow, power-law) decay of the first-order spatial coherence  function should apply in the disordered (quasi-ordered) phase, below (above) the critical point $P/P_{th}$, where $P_{th}$ is the pump power corresponding to the condensation threshold in our system. \cite{Berezinskii1973}. 
For each data sets of Fig.~\ref{fig:fig2}(a), we show the fitting {to Eqs.~\eqref{eq:gonefitrel}}  as solid blue and dashed red lines for the exponential and power-law fitting, respectively.
The power-law exponents $\alpha$ and the coherence lengths $\ell$ (extracted as fitting parameters) are depicted in Fig.~\ref{fig:fig2}(c),(d). At larger densities, $\ell$ becomes much larger than the trap size (dashed line) while $\alpha$ decreases asymptotically to zero, signaling full coherence across the system.

We recall that in the canonical BKT transition~\cite{Berezinskii1973}, the power-law decay of the coherence function is expected to exhibit a crossover in $\alpha$ from $\alpha_c = 0.25$ at the critical point towards $0$ for large densities. In the driven-dissipative case, it has been previously shown that the actual $\alpha_c$ may assume larger values due to the non-equilibrium nature of the polariton system introduced by the pump fluctuations~\cite{Roumpos2012PNAS,dagvadorj2015nonequilibrium}.
Aiming to identify the transition point, we proceed by comparing the root mean square (RMS) of the fitting curves, a method employed in a previous work to ascertain the phase of the system ~\cite{comaron2020BKT}. The analysis of the RMS of the fits (reported in the SI) reveals that the confined geometry introduces boundary effects resulting in a reduction of phase coherence. Moreover, even at very large densities, where power-law decay is expected, the exponential and power-law fits are still comparable. As a consequence, the ratio between the RMS of the fits cannot be used to identify the critical point~\cite{comaron2020BKT}.

%
\begin{figure*}[htp]
\centering
\includegraphics[width=\textwidth]{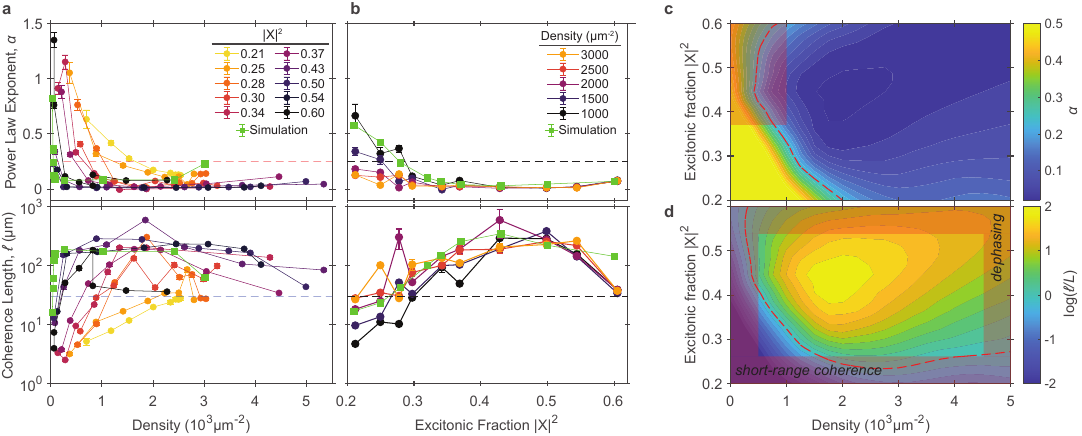}
\caption{\textbf{Dependence on excitonic fraction and polariton density}.
Extracted power-law exponents $\alpha$ (top) and coherence lengths $\ell$ (bottom), as a function of ({\bf a}) density and ({\bf b}) excitonic fraction, for different experimental series. Simulation results are presented as green squares.
{\bf c} Algebraic exponent $\alpha$ and {\bf d} coherence lenght $\ell$ as carpet plots in the density-excitonic-fraction plane. The highlighted region in {\bf c} is the region of interest in Fig.~\ref{fig:fig5}(f-i).
}
\label{fig:fig4}
\end{figure*}

We support our experimental finding with the numerical modeling of the polariton system using the stochastic driven-dissipative Gross-Pitaevskii equation~\cite{wouters2007}. As described in ``Methods'', we first evolve the model to reach a steady-state. Then, the first-order correlation function (see Methods) is numerically computed and plotted in Fig.~\ref{fig:fig3}(b) for different densities. The resulting outcomes are then fitted to exponential and power-law functions. Parameters from the fits are extracted and compared to the experimental values in Fig.~\ref{fig:fig2}(c,d).
Overall, a good qualitative agreement between the theory and the experiment is found, especially for larger pump powers (condensate densities).  

It is important to note that the slowest decay of correlations is attained, both in experiment and in simulations, only up to  $n\sim2\times10^3~\mathrm{\mu m^{-2}}$. For larger densities, we observe a loss of coherence, followed by an increase of $\alpha$ and a reduction of $\ell$.
The plot of the numerically computed pump power required to achieve a particular value of the condensate density, shown by a black solid line in Fig. ~\ref{fig:fig2}(d), confirms that the system needs a dramatically stronger pump power to reach larger densities. Strong external drive leads to a strong decoherence of the condensate.
{Pump sources, in fact, are expected to introduce noise contributions leading to dephasing at all powers \cite{ZHANG2022100399,Vallee92,Love2008}. At stronger powers we expect these to be large.
In addition, the dephasing  may also arise from ~\cite{Takemura2015} thermal fluctuations,  the incoherent excitonic reservoir~\cite{kavokin2017microcavities,Haug2010}, and polariton-polariton interactions~\cite{kim2016}.
 In our numerical simulations, the pump noise is explicitly included in the noise term $dW$ in Eq.~\eqref{eq:wigner} (see Methods), which accounts for the quantum fluctuations of the polariton field~\cite{carusotto2013quantum}
.}

\

\begin{figure*}[htp]
\centering
\includegraphics[width=\textwidth]{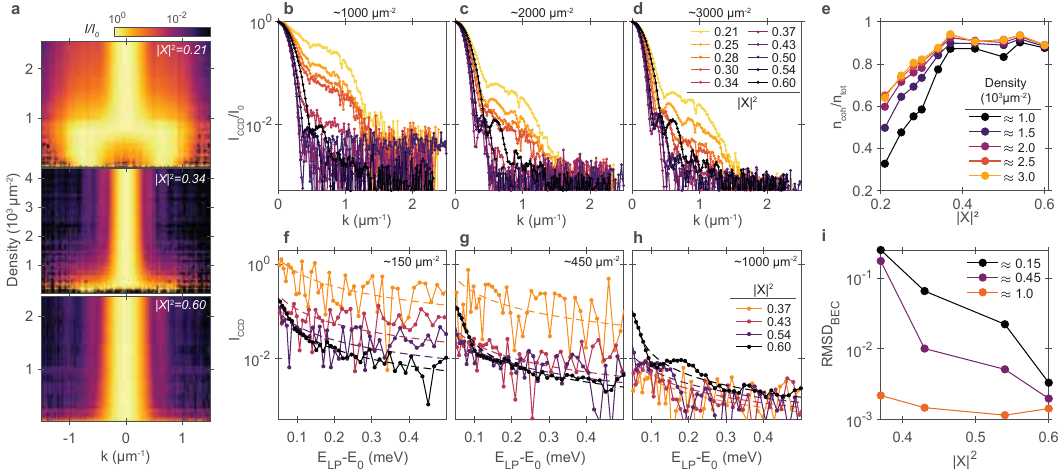}
\caption{\textbf{Thermalisation, coherent fraction, and dephasing}.
{\bf a} Momentum space distribution at different densities for three representative excitonic fractions $|X|^2 = 0.21$, $0.34$, $0.60$. 
The averaged-particle-number momentum distribution at different densities: $n \sim $ {\bf b} $10^3 \mathrm{\mu}m^{-2}$, {\bf c} $2 \times 10^3 \mathrm{\mu}m^{-2}$ and {\bf d} $3 \times 10^3 \mathrm{\mu}m^{-2}$. 
{\bf e} The coherent fraction extracted from the momentum distributions of polaritons.
{\bf f, g, h} At low densities the energy-distributions of polariton number (dots and solid lines) are fitted (dashed lines) with the Bose-Einstein distribution. RMS (root mean squares) of the fits are reported in  {\bf i}. 
}
\label{fig:fig5}
\end{figure*}

\noindent\textbf{{Dependence of coherence on the excitonic fraction at a fixed polariton density.} --}
To further understand the dependence of the spatial coherence on interactions, we change the excitonic fraction $|X|^2$ (i.e. the interaction strength, see Methods) and measure coherence for different polariton densities $n$.
We then investigate the behavior of spatial coherence as a function of $|X|^2$ for a fixed polariton density $n \sim 10^3 \mathrm{\mu m^{-2}}$.
We analyse the data for the coherence, similarly to the analysis presented in Fig.~\ref{fig:fig2}, and display the outcomes in Fig.~\ref{fig:fig3}.
Panel \ref{fig:fig3}(a) shows that the system undergoes a crossover from low to high coherence as the excitonic fraction of the polariton increases, quantified by the fitting curves and parameters shown in \ref{fig:fig3}(a) and (c),(d), respectively. This result clearly illustrates the interaction-driven nature of the coherence crossover for our system.
Interestingly, at large excitonic fractions, the system shows a lack of coherence that we attributed to the dephasing mechanisms observed at large pump powers and discussed in the previous section. In fact, to reach similar densities, larger excitonic fractions require stronger pumps. 
{This trend is clearly revealed by our simulations as shown in Fig.~\ref{fig:fig3}(d), which depicts the growth of the required pump power as a function of the excitonic fraction when the system density is kept constant.}
{Stronger  dephasing may be induced at larger excitonic fractions by stronger polariton-polariton and polariton-reservoir interactions, which increase with the excitonic Hopfield coefficient.
{Furthermore, the trapping potential becomes tighter at larger excitonic fraction, resulting in a stronger overlap of the condensate with the reservoir, as previously shown in similar experiments~\cite{estrecho2019direct}. The largest $|X|^2$ probed here already shows this effect.
}
}
{The experimental results shown in Fig.~\ref{fig:fig3}(a,c) and (d), are corroborated by the numerical simulations: the numerically extracted spatial coherence function, calculated at a fixed density ($n \sim 10^3 \mu m^{-2}$) for different values of $|X|^2$, are shown in Fig.~\ref{fig:fig3}(b). 
The extracted coherence lengths and power-law exponents are then compared to the experimental data in Fig.~\ref{fig:fig3} (c,d).
Remarkably, we find very good agreement between the theory and experiment in the limits of both small and large excitonic fractions.
}

Both increasing the excitonic fraction of the polaritons at a fixed density and increasing the density at a fixed excitonic fraction, leads to the growth of the interaction energy. To {better visualize} the coherence behaviour in the space of these two parameters, in Fig.~\ref{fig:fig4} we report the experimentally extracted $\alpha$ and $\ell$ as a function of density, Fig.~\ref{fig:fig4}(a), and as a function of excitonic fraction, Fig.~\ref{fig:fig4}(b). Panels (c) and (d) illustrate these parameters as a surface in the $(n,|X|^2)$ space, showing the onset of coherence at lengths above the system size (shown as a dashed red line).
In summary, in the confined system, coherence is enhanced when increasing the interaction energy by means of increasing either density or the excitonic fraction. The crossover to the full coherence takes place approximately at the system size, as predicted by the theory of trapped conservative quantum gases \cite{Dalfovo1999}.
{In the limits of both large densities and excitonic fractions, the polariton condensate experiences {dephasing mechanisms (due to the pump noise, polariton-polariton and polariton-exciton intercations)}, resulting in a reduction of coherence.
}

\

\noindent\textbf{{Coherent fraction and thermalisation.} --}
We further investigate possible relationship between the coherence and the thermalisation properties of our system.
We start by obtaining energy-resolved momentum-space imaging using a monochromator coupled with a CCD camera. 
The CCD intensity, $I_\mathrm{CCD}$, of the central section of each image is collected as a function of momenta, and recorded for different densities (pump powers) and excitonic fractions (exciton-photon detunings).
In Fig.~\ref{fig:fig5}(a) we report the normalised $I_\mathrm{CCD}$ for three cases: $|X|^2 = 0.21$, $0.34$, $0.60$.
At low $|X|^2$ and in the low density regime, the system is in a multimode regime. At large $|X|^2$, the system displays a single-mode occupation at all densities. This strongly suggests that a larger excitonic fraction leads to a more efficient energy relaxation towards the ground state~\cite{estrecho2019direct}.
In the context of canonical conservative condensates, quasi-long-range coherence leads to the formation of different coherent components~\cite{pitaevskii2003bose}: i) the condensate fraction, defined as the normalized density of particles which populate the zero-momentum $(k=0)$ mode, ii) the superfluid density defined \textit{\`a la} Nelson-and-Kosterlitz~\cite{Nelson1977} and iii) the quasicondensate, defined as a condensate with a fluctuating phase~\cite{kagan1987influence}. The superfluid density, in particular, can be directly linked to the power-law exponent $\alpha$ charactersising the BKT transition~\cite{Berezinskii1973,ProkofevSvistunov2002}.
For a non-equilibrium, open-dissipative condensate studied here, extraction of $n_s$ is diffcult. Instead, we introduce the notion of ``coherent density" $n_\mathrm{coh}$, defined as the fraction of particles below a certain momentum cut-off $k_\mathrm{cut}(|X|^2)$, the latter obtained by Fourier-transforming the effective trapping potential that the system experiences at each value of the excitonic fraction.

In order to extract the coherent fraction of the condensate, we first obtain the particle distribution per momentum state, $\mathcal{N}(k)$, which is proportional to the CCD intensity, i.e., $\mathcal{N}(k) = \mathcal{A} \times I_\mathrm{CCD}$ taking into account the correct geometry of the system~\cite{Pieczarka2019} {(see SI)}. Figures~\ref{fig:fig5}(b,c,d) show $I_\mathrm{CCD}$ for different $|X|^2$ at fixed densities $n=1$, $2$ and $3 \times 10^3 \mathrm{\mu m^{-2}}$.
From these distributions, we extract the coherent fraction $n_\mathrm{coh}/n_\mathrm{tot}$ in the (high-density) single-mode regime, and plot its behavior as a function of $|X|^2$ at different values of the total density $n_\mathrm{tot}$ in Figure~\ref{fig:fig5}(e). It is evident that the coherent fraction is increasing as a function of density and excitonic fraction (i.e., interaction strength). 
We note that a similar dependence on the interaction strength is expected for the superfluid fraction in a conservative system \cite{ProkofevSvistunov2002}.
For $|X|^2$ larger than $\sim 0.35$ the coherent fraction saturates at $n_\mathrm{coh}/n_\mathrm{tot} \sim 0.9$. 
At much larger excitonic fractions, the coherent fraction decreases, in agreement with the trend shown by the first-order correlation functions discussed in the previous sections. 

In the second part of this analysis, we discuss thermalisation properties of the system.
First, we move from momentum to energy space, employing the energy-momentum mapping provided by the dispersion relation of the lower-polariton branch $E_\mathrm{LP}(k) = 1/2 [ E_\mathrm{X}(k) +E_\mathrm{C}(k) -\sqrt{\hbar^2\Omega^2 +\delta^2}]$, where $E_\mathrm{X}(k)$ and $E_\mathrm{C}(k)$ are the energies of the exciton and cavity photon, respectively, $\delta(k)=E_\mathrm{C}-E_\mathrm{X}$ is the energy detuning, and $\hbar\Omega$ is the Rabi splitting energy.
We tentatively fit the data $\mathcal{N}(E)$ to the Bose-Einstein (BE) distribution of the average particle number $\mathcal{N}_\mathrm{BE}(E) = 1/(\exp((E-\mu)/k_B T)-1)$, where $T$ and $\mu$ are the temperature and chemical potential of the polaritons, respectively, and $k_B$ is the Boltzmann constant. Here we treat the temperature $T$ and chemical potential $\mu$ as free parameters. We are careful to apply this fitting only at low densities and high excitonic fraction, over a sufficiently populated tail of the modes which otherwise is obscured by a very bright (and peaked) main mode. This area is highlighted as a faint red shadowed region in Fig.~\ref{fig:fig4}(c)

In Fig.~\ref{fig:fig5}(f,g,h), we report the particle distributions in energy, with fitting curves as dashed lines. 
In panel (i) we show the RMS of the BE distributions; these give a quantitative tool for determining the goodness of the fit to a thermal distribution.
Our results suggest that the fitting to a thermal tail improves as the densities and excitonic fractions increase. 
Note that at larger densities ($n \sim 10^3 \mathrm{\mu m^2}$), the fitting procedure deteriorates for larger $|X|^2$; this is in connection with the increase of the $k=0$ peak which, at large excitonic fractions, dominates over the other modes.
Finally, we study the relation between the power-law exponent $\alpha$ and the coherent density $n_\mathrm{coh}$, with the aim to draw a parallel with the relationship between $\alpha$ and the superfuid density $n_s$ in conservative systems. In the canonical BKT theory, the two quantities are related by $\alpha = 1 /n_s \lambda^2$ where $\lambda = h / \sqrt{(2 \pi m k_BT)}$ is the thermal de-Broglie length \cite{Dalfovo1999}.

\begin{figure}[tp]
\centering
\includegraphics[width=\linewidth]{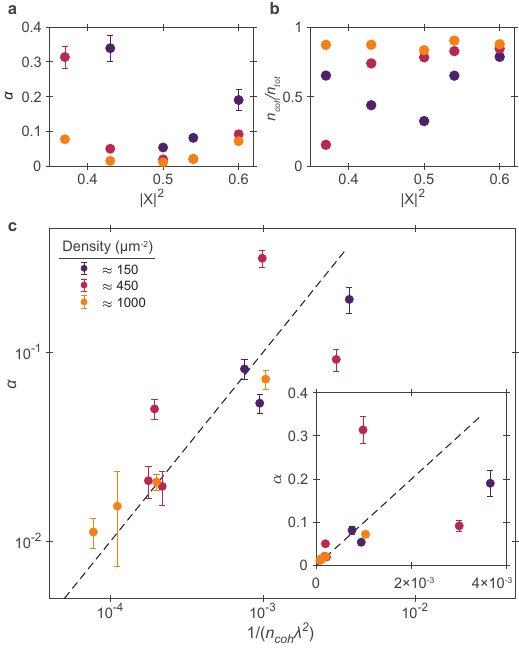}
\caption{\textbf{Power-law exponent and coherent fraction}.
 {\bf a} Algebraic exponents $\alpha$ and {\bf b} coherent fraction $n_\mathrm{coh}$ in the low-density and high excitonic fraction regime (highlighted in Fig.~\ref{fig:fig4} c,d as a dashed area), for different densities (see legend in {\bf c}).
{\bf c} Algebraic exponent shows an inverse proportionality to the coherent fraction, 
in analogy with the theory for conservative superfluids, where $\alpha = 1/n_\mathrm{s} \lambda^2$, where $\lambda = h/\sqrt{2 \pi m k_B T}$ is the thermal de-Broglie wavelength, and $n_\mathrm{s}$ is the superfluid density. 
{The dashed line corresponds to $\alpha = 10^2/(n_\mathrm{s} \lambda^2)$.}
Linear scales plot is shown in the inset.
}
\label{fig:fig6}
\end{figure}

The power-law exponent and the coherent fraction of the condensate in our system are plotted in Fig.~\ref{fig:fig6}(a,b), in the small-density and high excitonic component limit accessible from the previous analysis.
{Figure~\ref{fig:fig6}(c) clearly depicts the linear dependence $\alpha = 1/{n_\mathrm{coh}}\lambda^2$ for the data sets obtained in our experiment. Note that this linear behaviour persists across almost two orders of magnitude for both $\alpha$ and $1/{n_\mathrm{coh}}\lambda^2$. It also shows a proportionality constant diverging from the expected value of 1 by two orders of magnitude, {a deviation} that may be due to the driven-dissipative and/or confined nature of the system.}

\

\noindent\textbf{\large {Discussion} \normalsize}

In this work, we investigate the role of interactions and energy relaxation on the development of spatial coherence in the confined polariton system.
Our experimental and numerical findings confirm that, in the driven-dissipative scenario of polariton condensation, the system follows a BKT-type phase transition, which is driven by interactions. 
As the interaction energy is tuned across a wide range (up to 700~$\mu$eV) by varying the density and interparticle interactions, we observe different regimes of coherence. In the high-density regime of stronger interacting, highly excitonic polaritons full coherence across the spatial size of the system can be reached. 
However, the coherence is degraded above certain densities and excitonic fractions. {Our results suggest that this behavior is driven by the dephasing due to the fluctuations of the laser pump,  which are stronger at higher pump powers, as well as due to polariton-polariton interactions and interactions of polaritons with the reservoir excitons. This is exacerbated in the experiments due to the tightening of the reservoir-induced trap at higher excitonic fractions~\cite{estrecho2019direct}, resulting in stronger reservoir-condensate overlap.}

In the low-density regime, we clearly see the short-range thermal component of the first-order correlation function~\cite{KrugerPRL2007}.
In the limit of highly photonic polaritons, we observe multi-mode condensation due to inefficient energy relaxation towards the ground state. 
We find that the fitting to BE statistics improves for stronger interactions and at larger densities, proving that thermalisation processes in non-equilibrium condensate take place via interactions. 
We note that this inhibited energy relaxation towards the ground state precludes the observation of a BEC phase in confined, 2D gas of weakly interacting, highly photonic polaritons.

{Finally, we tentatively investigate the relationship between the power-law exponent of $g^{(1)}(r)$, the condensate coherent fraction, and the effective temperature of the system. 
Remarkably, we find that the power-law exponent is directly proportional to the temperature of the system, which defines the thermal de Broglie wavelength of a polariton, and is  inversely proportional to the coherent density, in analogy to the dependence of the power-law exponent on the superfluid density in the conservative case. 
However, we note that the system temperature in conservative systems is extracted from the fitting with a thermal distribution, while in a driven-dissipative system, full equilibration is not guaranteed and the BE fitting is useful only in a limited subset of the system's parameters. 
To our knowledge, prior to this study, no investigation had explored this particular relation in the context of nonequilibrium, driven-dissipative quantum systems. Further experimental effort could address the issue of the extraction of the superfluid density and its relation to the coherent density defined here.

{Summarising, our} work represents an important step towards the characterisation of interaction-mediated BKT phase transition and coherence in the driven-dissipative quantum system, across a range of interaction energy. Our findings are particularly relevant to the discussion about superfluid properties~\cite{juggins2018}, universal behaviour~\cite{Fontaine2022}, and turbulent features~\cite{panico2023} in driven-dissipative quantum systems.

\

\noindent \textbf{\large {Methods}  \normalsize}

\noindent\textbf{{Sample and experimental setup.}}
We use a high Q-factor GaAs/AlGaAs planar microcavity with 7-nm GaAs quantum wells embedded in a 3$\lambda$/2-cavity, similar to the samples used in Refs.~\cite{steger2015slow, caputo2018topological}. The thickness gradient of the sample enables control of the exciton-photon detuning by probing different positions on the sample. The 720-nm off-resonant continuous-wave (CW) excitation laser is spatially structured into a ring profile using an axicon, a pair of lenses, and a 50X 0.5-NA objective. The same objective collects the polariton emission which goes through a relay of lenses for real-space imaging with a SCMOS camera (Andor Zyla 4.2). The camera images the interference fringes used to measure the first-order correlation function. A monochromator (Princeton Instruments IsoPlane SCT 320) coupled with a CCD camera (Andor iXon Ultra 888) enables energy-resolved real and momentum-space imaging.

\noindent\textbf{{Tuning the excitonic fraction.}}
The excitonic fraction $|X|^2$ is tuned by probing areas of the samples with different photon-exciton detuning, $\Delta$. The sample has a gradient in detuning due to a wedge in the effective cavity length~\cite{steger2015slow} resulting in a gradient in cavity photon energy. The detuning is calculated from the polariton dispersion, measured at very low excitation power (below condensation threshold) for each position on the sample, and using previously known parameters~\cite{estrecho2019direct,pieczarka2020observation}. We then extract the excitonic fraction from the detuning as $|X|^2 = 1/2\left( 1+\Delta/\sqrt{\Delta^2 + \Omega^2}\right)$, where the Rabi splitting is about $\Omega=15.9$~meV~\cite{pieczarka2020observation}.

\noindent\textbf{{Coherence measurement setup.}}
We employ a modified interferometer~\cite{caputo2018topological} consisting of a 50:50 beam splitter, a flat mirror, and a hollow retroreflector to measure the spatial first-order correlation function $|g^{(1)}|$. The setup combines the polariton real-space emission $I(\mathbf{r})=|\psi(\mathbf{r})|^2$ with its inverted image $I(-\mathbf{r})=|\psi(-\mathbf{r})|^2$ onto a sCMOS camera where the interferogram image $F(\mathbf{r})$ forms. The mirrors are aligned such that the autocorrelation point, $(\Delta x, \Delta y))=0$, is at the center of the real-space distribution. The first-order correlation function is then extracted using the equation
$$|g^{(1)}(\mathbf{r},-\mathbf{r};\Delta t=0)|=V(\mathbf{r})\frac{I(\mathbf{r}) + I(-\mathbf{r})}{2\sqrt{I(\mathbf{r}) I(-\mathbf{r})}} $$
where $V = {(F_{max} - F_{min})}/{(F_{max} + F_{min})}$ is the visibility of the interference fringes extracted by introducing sub-micron delays using a piezo actuator~\cite{caputo2018topological}.
The mirrors are mounted on motorized stages to ensure near-zero temporal delay $\Delta t\approx 0$. A piezo actuator provides a series of sub-wavelength delays near $\Delta t\approx 0$ to accurately determine the spatial variation of the fringe visibility, $V(\mathbf{r})$.
To filter out low-frequency noise coming from the mechanical vibrations of the whole experimental setup, the CW laser is chopped down to $\sim$10-100-$\mu s$ pulse lengths depending on the signal-to-noise ratio. This time scale is orders of magnitude longer than the polariton lifetime ($\sim$100-ps) and the pulse rise time ($\sim$50 ns), and hence is sufficient to capture the steady-state regime.

\noindent\textbf{{Modelling of the polariton field.}}
We describe the effective dynamics of the polariton fluid through the equation of motions for the two-dimensional polariton field  $\psi=\psi(\bm{r},t)$ as a function of the position $\bm{r}=(x,y)$ and time $t$ within the truncated Wigner approximation\cite{carusotto2013quantum}, which reads ($\hbar=1$)~\cite{wouters2007,chiocchetta2013,comaron2018dynamical}:
\begin{multline}
id \psi = dt\bigg[ \left( i \beta - 1 \right) \frac{\nabla^2}{2 m_{pol}} + g_c|{\psi}|^2_{-} + \\
+ \frac{i }{2} \frac{R}{\gamma_R}
\bigg( \frac{P(\textbf{r})}{1+\frac{R}{\gamma_R}|{\psi}|^2_{-}} -
\gamma \bigg) \bigg]
\psi +  dW
\label{eq:wigner}
\end{multline}
where $m$ is the polariton mass, $\gamma_c$ and $\gamma_R$ are the polariton and excitonic reservoir decay rates, $P(\textbf{r})$ the circularly-shaped external drive, $g_c$ is the polariton-polariton interaction strength. $R$ is the scattering rate of reservoir particles into the condensate.
The renormalised density $|{\psi}|^2_{-} \equiv
\left(\left|{\psi} \right|^2 - {1}/{2dV} \right)$  includes the subtraction of the Wigner commutator contribution (where $dV=a^2$ is the element of volume of our 2D grid with the lattice spacing $a$).
The zero-mean white Wigner noise $dW$ fulfils the condition $\langle
dW^{*}(\vec{r},t) dW (\vec{r}',t) \rangle
=[((R / \gamma_R) P +\gamma_c)/2] \delta_{\vec{r},\vec{r}'}dt$.
The equation above includes the phenomenological term proportional to the constant $\beta$, which quantifies the rate of energy relaxation in the system~\cite{chiocchetta2013,woutersLiew2010,wouters2010,comaron2018dynamical,zamora2020}.
The model Eq.~\eqref{eq:wigner} corresponds to the adiabatic approximation limit of the generalised polariton equations of motion coupled to an external reservoir~\cite{wouters2007}, which is justified once the reservoir is able to adiabatically follow the evolution of the condensate and $\gamma_c \ll \gamma_R$~\cite{bobrovska2015}.

Using the parameters adopted in Ref.~\cite{estrecho2018single} to model an experimental system similar to that described here, in this work we vary the excitonic fraction $|X|^2$ in order to model the steady-state coherence across the range of interaction strengths. It is therefore crucial to note that most of the parameters in Eq.~\eqref{eq:wigner} depend on the excitonic Hopfield coefficient $X$. The latter also defines the value of the excitonic fraction $|X|^2$, and depends on the exciton–photon detuning $\Delta$ and the Rabi splitting $\hbar\Omega$ as $|X|^2~=~{1}/{2}\left(1+ {\Delta}/{\sqrt{4 \hbar^2 \Omega^2 + \Delta^2}}, \right)$.
The polariton decay rate depends on the excitonic fraction as  $\gamma_c = (1-|X|^2) \gamma_{ph}$, where $\gamma_{ph} = 1/\tau_{ph}$ corresponds to the inverse of the photon lifetime.
The constants $g$ and $g_R$ characterise the strengths of polariton-polariton and polariton-reservoir interactions, which become stronger for polaritons with a larger excitonic fraction. They can be estimated $g_c = g_{ex}|X|^4$, $g_R = g_{ex}|X|^2$~\cite{Bleu2020}, with $g_{ex}$ the exciton-exciton interaction. 
Following the discussion in Ref. ~\cite{estrecho2018single}, we assume that the stimulated scattering rate $R$ from the reservoir into the polariton states is more efficient for more excitonic polaritons: from our results we find that $R = R_0 (g_c/g_R)^2$, namely $R$ scales quadratically with $|X|^2$.

The circularly symmetric pump profile $P(\textbf{r})$ is implemented as the sum of two contributions: i) the  profile $P(\textbf{r}) = P_0 {\gamma_c}/{R} [ e^{-\left( {\mathcal{E}}/{2 \sigma^2}\right)^2}]$, where $P_0$ is the pump strength, and $\mathcal{E} = x^2/a^2 +y^2/a^2 -1$ is the circle equation with radius $R_P$, and ii) a weak flat pump $\bar{P} = \alpha_p P_0 \gamma_c/R$, with $\alpha_p < 1$ positive and small, at the center of the ellipse for $\textbf{r} < R_P$: with the latter, we account for the presence of a non-vanishing reservoir density at the center of the trap.
We note that the inclusion of such a term is important for enhancing the onset of coherence at the centre of the trap.
We also find that the experimental data are adequately modeled by relaxation mechanisms~\cite{wouters2010,Wouters_2012,woutersLiew2010} (controlled by the paramater $\beta)$
being more efficient for larger excitonic fractions as $\beta = \beta_0 + \beta_1 |X|^2$.
In order to match the experimental measurements, we use the following parameters: 
$m=3.6 \times 10^{-5} m_e$ with $m_e$ the electron mass, $\tau_{ph} = 135 \si{ps}$, 
$\gamma_R = 10^{-3} \mathrm{ps}^{-1}$, 
$g_{ex}~=1.12~\si{\mu eV \mu m^2}$, 
$R_0 = 1.4 \times 10^{-3} \si{\mu m^2 ps^{-1}}$. 
The parameters of the pump ring read: the radius $R_P = R_{in} + (R_{out}-R_{in})/2$, with $R_{out} = 15 \si{\mu m}$ and $R_{in} = 13 \si{\mu m}$, and $\sigma_{ring}= 0.63 \si{\mu m}$, corresponding to a FWHM of the laser of $1.5 \si{\mu m}$. For the relaxation coefficients we use $\beta_0 \approx 0.1$ and $\beta_1 \approx 0.5$. The weak flat pump at the center of the trap has an amplitude parameter $\alpha_p = 0.1$.
We simulate the dynamics of the polariton system by numerically integrating the stochastic differential equations Eq.~\eqref{eq:wigner} for the polariton field.
The numerical integration is performed on a two-dimensional lattice with Dirichlet boundary conditions. The lattice is composed of $128^2$ grid points, with the lattice spacing $\Delta x = 0.63\mu m$. 
The lattice spacing is chosen to lie within the validity regime of the truncated Wigner methods used for the description of the stochastic field equations~\cite{carusotto2013quantum,comaron2018dynamical}, but large enough to capture the macroscopic physics of the system. We notice that it also introduces a cut-off $k_{cut} \propto \Delta x^{-1}$ in the momentum representation of the field.
The integration of Eq.~\eqref{eq:wigner} in time is performed by using the XMDS2 software framework~\cite{dennis2013xmds2}. Specifically, we use a fixed time step which ensures stochastic noise consistency, and a fourth-order Runge-Kutta algorithm. 
Once the steady-state is achieved, to obtain a smooth first-order correlation function, $g^{(1)}(r)$, we average over both the correlators at the $x=0$ and $y=0$ positions, using the expression $ g^{(1)}(r) \equiv [ g_x^{(1)}(r) + g_y^{(1)}(r)]/2 $ where 
$g_x^{(1)}(r) = 
{\langle\psi^*_{i,0}\psi_{i+r,0}\rangle_{\mathcal{N}}}/{\sqrt{\langle |\psi_{i,0}|^2\rangle_{\mathcal{N}} \langle |\psi_{i+r,0}|^2\rangle_{\mathcal{N}}}}$ and $
g_y^{(1)}(r) = 
{\langle\psi^*_{0,j}\psi_{0,j+r}\rangle_{\mathcal{N}}}/{\sqrt{\langle |\psi_{0,j}|^2\rangle_{\mathcal{N}} \langle |\psi_{0,j+r}|^2\rangle_{\mathcal{N}}}} $.
Here, $\psi_{i,j} = \psi(x_i,y_j)$ and the average $\langle \dots \rangle_\mathcal{N}$ is performed over the number $\mathcal{N}$ of stochastic realisations. 
All the results presented in our study are converged with respect to the number of stochastic realisations $\mathcal{N
} = 100$.

\


\bigskip
\noindent\textbf{\large Acknowledgements \normalsize}\\ 
We thank H. Weinberger for fruitful discussions and for proofreading the text.
This research made use of the HPC Computing service at University College of London. We thank F. Ducluzeau and F. Sidoli for computational assistance.
PC and MSW acknowledge financial support from EPSRC (Grants No. EP/S019669/1, and No. EP/V026496/1). MM acknowledges support from the National Science Center, Poland grant No.~2021/43/B/ST3/00752. EE, MW, and EAO acknowledge support from the Australian Research Council (ARC) through the Centre of Excellence Grant CE170100039. 
EE acknowledges support from the ARC Discovery Early Career Researcher Award (DE220100712). MW acknowledges support by Schmidt Science Fellows, in partnership with Rhodes Trust.
\\

\noindent\textbf{\large Additional information \normalsize}\\ 
\textbf{Supplementary Information} accompanies this paper at (link)\\

\noindent\textbf{Competing interests:} The authors declare no competing interests.\\

\end{document}